\documentclass[aps,prl,twocolumn,superscriptaddress,showpacs,letterpaper]{revtex4}

\usepackage{graphicx}
\usepackage{amsmath, amsthm, amssymb}

\def\cdf2{CDF II}
\newcommand{\fb}{\ifmmode  {\rm fb}^{-1}
                 \else
                           \mbox{fb$^{-1}$}
                 \fi
                }
\newcommand{\mevcc}{\ifmmode {\rm MeV}/c^2%
                 \else%
                           \mbox{MeV}$/c^2$%
                 \fi%
                 }%
\newcommand{\ra}{\rightarrow}
\newcommand{\pt}{\ifmmode  p_{\rm T}%
                 \else%
                            \mbox{$p_{\rm{T}}$ }%
                 \fi%
                 }%

\begin{document}

\title{Measurement of Resonance Parameters of
	Orbitally Excited Narrow $B^0$ Mesons}

\affiliation{Institute of Physics, Academia Sinica, Taipei, Taiwan 11529, Republic of China} 
\affiliation{Argonne National Laboratory, Argonne, Illinois 60439} 
\affiliation{University of Athens, 157 71 Athens, Greece} 
\affiliation{Institut de Fisica d'Altes Energies, Universitat Autonoma de Barcelona, E-08193, Bellaterra (Barcelona), Spain} 
\affiliation{Baylor University, Waco, Texas  76798} 
\affiliation{Istituto Nazionale di Fisica Nucleare Bologna, $^w$University of Bologna, I-40127 Bologna, Italy} 
\affiliation{Brandeis University, Waltham, Massachusetts 02254} 
\affiliation{University of California, Davis, Davis, California  95616} 
\affiliation{University of California, Los Angeles, Los Angeles, California  90024} 
\affiliation{University of California, San Diego, La Jolla, California  92093} 
\affiliation{University of California, Santa Barbara, Santa Barbara, California 93106} 
\affiliation{Instituto de Fisica de Cantabria, CSIC-University of Cantabria, 39005 Santander, Spain} 
\affiliation{Carnegie Mellon University, Pittsburgh, PA  15213} 
\affiliation{Enrico Fermi Institute, University of Chicago, Chicago, Illinois 60637} 
\affiliation{Comenius University, 842 48 Bratislava, Slovakia; Institute of Experimental Physics, 040 01 Kosice, Slovakia} 
\affiliation{Joint Institute for Nuclear Research, RU-141980 Dubna, Russia} 
\affiliation{Duke University, Durham, North Carolina  27708} 
\affiliation{Fermi National Accelerator Laboratory, Batavia, Illinois 60510} 
\affiliation{University of Florida, Gainesville, Florida  32611} 
\affiliation{Laboratori Nazionali di Frascati, Istituto Nazionale di Fisica Nucleare, I-00044 Frascati, Italy} 
\affiliation{University of Geneva, CH-1211 Geneva 4, Switzerland} 
\affiliation{Glasgow University, Glasgow G12 8QQ, United Kingdom} 
\affiliation{Harvard University, Cambridge, Massachusetts 02138} 
\affiliation{Division of High Energy Physics, Department of Physics, University of Helsinki and Helsinki Institute of Physics, FIN-00014, Helsinki, Finland} 
\affiliation{University of Illinois, Urbana, Illinois 61801} 
\affiliation{The Johns Hopkins University, Baltimore, Maryland 21218} 
\affiliation{Institut f\"{u}r Experimentelle Kernphysik, Universit\"{a}t Karlsruhe, 76128 Karlsruhe, Germany} 
\affiliation{Center for High Energy Physics: Kyungpook National University, Daegu 702-701, Korea; Seoul National University, Seoul 151-742, Korea; Sungkyunkwan University, Suwon 440-746, Korea; Korea Institute of Science and Technology Information, Daejeon, 305-806, Korea; Chonnam National University, Gwangju, 500-757, Korea} 
\affiliation{Ernest Orlando Lawrence Berkeley National Laboratory, Berkeley, California 94720} 
\affiliation{University of Liverpool, Liverpool L69 7ZE, United Kingdom} 
\affiliation{University College London, London WC1E 6BT, United Kingdom} 
\affiliation{Centro de Investigaciones Energeticas Medioambientales y Tecnologicas, E-28040 Madrid, Spain} 
\affiliation{Massachusetts Institute of Technology, Cambridge, Massachusetts  02139} 
\affiliation{Institute of Particle Physics: McGill University, Montr\'{e}al, Canada H3A~2T8; and University of Toronto, Toronto, Canada M5S~1A7} 
\affiliation{University of Michigan, Ann Arbor, Michigan 48109} 
\affiliation{Michigan State University, East Lansing, Michigan  48824}
\affiliation{Institution for Theoretical and Experimental Physics, ITEP, Moscow 117259, Russia} 
\affiliation{University of New Mexico, Albuquerque, New Mexico 87131} 
\affiliation{Northwestern University, Evanston, Illinois  60208} 
\affiliation{The Ohio State University, Columbus, Ohio  43210} 
\affiliation{Okayama University, Okayama 700-8530, Japan} 
\affiliation{Osaka City University, Osaka 588, Japan} 
\affiliation{University of Oxford, Oxford OX1 3RH, United Kingdom} 
\affiliation{Istituto Nazionale di Fisica Nucleare, Sezione di Padova-Trento, $^x$University of Padova, I-35131 Padova, Italy} 
\affiliation{LPNHE, Universite Pierre et Marie Curie/IN2P3-CNRS, UMR7585, Paris, F-75252 France} 
\affiliation{University of Pennsylvania, Philadelphia, Pennsylvania 19104}
\affiliation{Istituto Nazionale di Fisica Nucleare Pisa, $^y$University of Pisa, $^z$University of Siena and $^{aa}$Scuola Normale Superiore, I-56127 Pisa, Italy} 
\affiliation{University of Pittsburgh, Pittsburgh, Pennsylvania 15260} 
\affiliation{Purdue University, West Lafayette, Indiana 47907} 
\affiliation{University of Rochester, Rochester, New York 14627} 
\affiliation{The Rockefeller University, New York, New York 10021} 
\affiliation{Istituto Nazionale di Fisica Nucleare, Sezione di Roma 1, $^{bb}$Sapienza Universit\`{a} di Roma, I-00185 Roma, Italy} 

\affiliation{Rutgers University, Piscataway, New Jersey 08855} 
\affiliation{Texas A\&M University, College Station, Texas 77843} 
\affiliation{Istituto Nazionale di Fisica Nucleare Trieste/Udine, $^{cc}$University of Trieste/Udine, Italy} 
\affiliation{University of Tsukuba, Tsukuba, Ibaraki 305, Japan} 
\affiliation{Tufts University, Medford, Massachusetts 02155} 
\affiliation{Waseda University, Tokyo 169, Japan} 
\affiliation{Wayne State University, Detroit, Michigan  48201} 
\affiliation{University of Wisconsin, Madison, Wisconsin 53706} 
\affiliation{Yale University, New Haven, Connecticut 06520} 
\author{T.~Aaltonen}
\affiliation{Division of High Energy Physics, Department of Physics, University of Helsinki and Helsinki Institute of Physics, FIN-00014, Helsinki, Finland}
\author{J.~Adelman}
\affiliation{Enrico Fermi Institute, University of Chicago, Chicago, Illinois 60637}
\author{T.~Akimoto}
\affiliation{University of Tsukuba, Tsukuba, Ibaraki 305, Japan}
\author{M.G.~Albrow}
\affiliation{Fermi National Accelerator Laboratory, Batavia, Illinois 60510}
\author{B.~\'{A}lvarez~Gonz\'{a}lez}
\affiliation{Instituto de Fisica de Cantabria, CSIC-University of Cantabria, 39005 Santander, Spain}
\author{S.~Amerio$^x$}
\affiliation{Istituto Nazionale di Fisica Nucleare, Sezione di Padova-Trento, $^x$University of Padova, I-35131 Padova, Italy} 

\author{D.~Amidei}
\affiliation{University of Michigan, Ann Arbor, Michigan 48109}
\author{A.~Anastassov}
\affiliation{Northwestern University, Evanston, Illinois  60208}
\author{A.~Annovi}
\affiliation{Laboratori Nazionali di Frascati, Istituto Nazionale di Fisica Nucleare, I-00044 Frascati, Italy}
\author{J.~Antos}
\affiliation{Comenius University, 842 48 Bratislava, Slovakia; Institute of Experimental Physics, 040 01 Kosice, Slovakia}
% ...

\collaboration{CDF Collaboration\footnote{With visitors from $^a$University of Massachusetts Amherst, Amherst, Massachusetts 01003,
$^b$Universiteit Antwerpen, B-2610 Antwerp, Belgium, 
$^c$University of Bristol, Bristol BS8 1TL, United Kingdom,
$^d$Chinese Academy of Sciences, Beijing 100864, China, 
$^e$Istituto Nazionale di Fisica Nucleare, Sezione di Cagliari, 09042 Monserrato (Cagliari), Italy,
$^f$University of California Irvine, Irvine, CA  92697, 
$^g$University of California Santa Cruz, Santa Cruz, CA  95064, 
$^h$Cornell University, Ithaca, NY  14853, 
$^i$University of Cyprus, Nicosia CY-1678, Cyprus, 
$^j$University College Dublin, Dublin 4, Ireland,
$^k$Royal Society of Edinburgh/Scottish Executive Support Research Fellow,
$^l$University of Edinburgh, Edinburgh EH9 3JZ, United Kingdom, 
$^m$Universidad Iberoamericana, Mexico D.F., Mexico,
$^n$Queen Mary, University of London, London, E1 4NS, England,
$^o$University of Manchester, Manchester M13 9PL, England, 
$^p$Nagasaki Institute of Applied Science, Nagasaki, Japan, 
$^q$University of Notre Dame, Notre Dame, IN 46556,
$^r$University de Oviedo, E-33007 Oviedo, Spain, 
$^s$Simon Fraser University, Vancouver, British Columbia, Canada V6B 5K3,
$^t$Texas Tech University, Lubbock, TX  79409, 
$^u$IFIC(CSIC-Universitat de Valencia), 46071 Valencia, Spain,
$^v$University of Virginia, Charlottesville, VA  22904,
$^{dd}$On leave from J.~Stefan Institute, Ljubljana, Slovenia, 
}}
\noaffiliation


\begin{abstract}
We report a measurement of resonance parameters of the
orbitally excited ($L=1$) narrow $B^0$ mesons in decays
to $B^{(*)+}\pi^-$ using $1.7$ \fb of data collected
by the \cdf2 detector at the Fermilab Tevatron. 
The mass and width of the $B^{*0}_2$ state are 
measured to be 
$m(B^{*0}_2) = 5740.2^{+1.7}_{-1.8}$(stat.)$^{+0.9}_{-0.8}$(syst.) MeV/$c^2$
and 
$\Gamma(B^{*0}_2) = 22.7^{+3.8}_{-3.2}$(stat.)$^{+3.2}_{-10.2}$(syst.) MeV/$c^2$.
The mass difference between the $B^{*0}_2$
and $B^0_1$ states is measured to be 
$14.9^{+2.2}_{-2.5}$(stat.)$^{+1.2}_{-1.4}$(syst.) MeV/$c^2$, 
resulting in a $B^0_1$ mass of 
$5725.3^{+1.6}_{-2.2}$(stat.)$^{+1.4}_{-1.5}$(syst.) MeV/$c^2$.
This is currently the most precise measurement of the masses of
these states and the first measurement of
the $B^{*0}_2$ width.
\end{abstract}

\maketitle

Mesons consisting of a light and a heavy quark are an interesting 
laboratory for the study of quantum chromodynamics, the theory of 
strong interactions.  The role of the heavy-light quark mesons is
similar to that played by the hydrogen atom in understanding 
quantum electrodynamics.  The bound states of a $\bar{b}$ quark 
with either a light $u$ or $d$ quark are referred to as $B$ mesons.
The states with zero internal orbital angular momentum ($L=0$) and 
spin parity $J^P = 0^-$ ($B$) and $1^-$ ($B^*$) are well established
\cite{pdg}, but the spectroscopy of the orbitally excited $B$ states 
has not been well studied.
For $L=1$, the total angular momentum of the light quark is 
$j = \frac{1}{2}$ or $j = \frac{3}{2}$.  With the addition of
spin of the heavy quark, two doublets of states are expected: 
states with $j = \frac{1}{2}$, named $B^*_0$ ($J=0$) and $B_1^\prime$ ($J=1$),
and states with $j = \frac{3}{2}$, named $B_1$ ($J=1$) and $B^*_2$ ($J=2$).
These four states are collectively referred to as $B^{**}$.

Heavy quark effective theory~\cite{hqet} predicts that the 
mass splitting within each doublet of a heavy-light quark meson
is inversely proportional to the heavy quark mass
\cite{hqet,Eichten:1993ub,Falk:1995th,Isgur:1998kr,Ebert:1997nk,Orsland:1998de,Matsuki:2007zz}.
The $j = \frac{1}{2}$ states are expected to decay to $B^{(*)}\pi$
via an S-wave transition and to exhibit resonance widths in the range 
$100-200~\mevcc$~\cite{TMehen}.  The $j = \frac{3}{2}$ states are 
expected to decay to $B^{(*)} \pi$ via a D-wave transition and to have 
widths of $10-20~\mevcc$~\cite{Falk:1995th,Orsland:1998de}.
This Letter focuses on the $B_1$ and $B^*_2$ observed in $B\pi$ final 
states.  The decay $B_1 \to B\pi$ is forbidden by conservation of angular 
momentum and parity, while both $B^*_2 \to B\pi$ and $B^*_2 \to B^*\pi$ 
decays are allowed. Decays to a $B^*$ are followed by $B^*\rightarrow B\gamma$, 
where the photon is not reconstructed in CDF due to its low energy.  
Because of the missing photon, the measured $B\pi$ mass in 
$B_1\to B^*\pi \to B\pi\gamma$ and $B_2^*\to B^*\pi \to B\pi\gamma$ 
events is lower than the $B^*\pi$ mass by $45.78\pm0.35~\mevcc$~\cite{pdg},
resulting in an expected signal structure of three narrow 
$B\pi$ peaks for the $B_1$ and $B^{*}_2$.

Previous measurements of properties of the $j=\frac{3}{2}$ $B_1^0$ and 
$B^{*0}_2$ mesons using inclusive or partially reconstructed decays 
did not separate the narrow states 
\cite{previous-measurements,delphi_bds_ichep}
or were limited by low sample statistics~\cite{Barate:1998cq}.
Recently the D0 Collaboration resolved the 
$B_1^0$ and $B^{*0}_2$ masses~\cite{Abazov:2007vq}. 
The superb mass resolution of the \cdf2 detector allows better 
precision and enables us to measure the $B^{*0}_2$ width.
%
Here, we present measurements of the masses of the $B^0_1$ and 
$B^{*0}_2$ states and the width of the $B_2^{*0}$ state. We 
reconstruct $B^{**0}$ in $B^{+}\pi^-$ and $B^{*+}\pi^-$ decays, 
where the $B^+$ candidates decay into $J/\psi K^+$, 
$\bar{D}^0\pi^+$, and $\bar{D}^0\pi^+\pi^+\pi^-$ final states with
$J/\psi\ra\mu^+\mu^-$ and $\bar{D}^0\ra K^+\pi^-$.
Throughout this paper, any reference to a
specific charge state implies the charge conjugate state as well.

We use a data sample of events produced in $p\bar{p}$ collisions 
at $\sqrt{s} = 1.96$~TeV recorded by the \cdf2 detector at the Tevatron,
corresponding to an integrated luminosity of $1.7~\mathrm{fb^{-1}}$.  
The components and performance parameters of \cdf2 \cite{Acosta:2004yw} 
most relevant for this analysis are the tracking,
the muon detectors, and the trigger on displaced vertices.
The tracking system lies in a uniform axial magnetic field
of $1.4$~T.  The inner tracking volume is instrumented with a 
layer of single-sided silicon microstrip detectors mounted 
directly on the beam pipe at a  radius of $1.5$~cm, and $7$ 
layers of double-sided silicon that extend out to a radius of 
$28$~cm~\cite{Hill:2004qb}.  This system provides excellent 
resolution of the impact parameter, $d_0$, defined as the 
distance of closest approach of the track to the interaction 
point in the transverse plane.
The outer tracking volume contains an open-cell drift chamber 
(COT) up to a radius of $137$~cm \cite{Affolder:2003ep}.
Muons are detected in planes of drift tubes and scintillators
\cite{Ascoli:1987av} located outside the hadronic and
electromagnetic calorimeters.  The muon detectors used in this study
cover the pseudorapidity range $\mathrm{|\eta|\le 1.0}$, where 
$\eta=-\ln\tan(\theta/2)$ and $\theta$ is the polar angle measured 
from the proton beam.

A three-level trigger system selects events in real time.
A dimuon trigger \cite{Acosta:2004yw} requires two tracks 
of opposite charge that match track segments in the muon
chambers and have a combined dimuon mass consistent with 
the $J/\psi$ mass.
An extremely fast tracker at level 1 (XFT)~\cite{XFT}
groups COT hits into tracks in the transverse plane. 
A silicon vertex trigger at level 2 (SVT)~\cite{SVT}
adds silicon hits to tracks found by the XFT, thus
providing better-defined tracks and allowing candidate 
selection based on the impact parameter.
A displaced vertex trigger~\cite{DTrig} requires two 
tracks each with a scalar transverse momentum, $p_T$, greater 
than $2$~GeV$/c$ and with $0.12 < d_0 < 1~\mbox{mm}$.
Additionally, the intersection point of the track pair must be 
transversely displaced from the $p\bar{p}$ interaction point by 
at least $0.2~\mbox{mm}$, and the pair must have a scalar sum 
$p_T(1)+p_T(2) > 5.5$~GeV$/c$.

Decays $B^+\ra J/\psi K^+$ are reconstructed from the dimuon 
trigger data while decays $B^+\ra \bar{D}^0\pi^+(\pi^+\pi^-)$
are reconstructed from the displaced vertex trigger data. 
In each decay, the tracks are constrained in a three-dimensional 
kinematic fit to the appropriate $B^+$ vertex topology with the 
$J/\psi$ and $\bar{D}^0$ masses constrained to the world average 
values~\cite{pdg}.  Each track compatible with originating from
the same interaction point as the $B^+$ and not used to reconstruct 
the $B^+$ is considered as a pion candidate, and its 4-momentum is 
combined with that of the $B^+$ candidate to form a $B^{**0}$ candidate.
We search for narrow resonances in the mass dif\mbox{}ference 
distribution of $Q = m(B^+\pi^-)-m(B^+)-m_\pi$, where $m(B^+\pi^-)$ 
and $m(B^+)$ are the reconstructed invariant masses of the $B^+\,\pi^-$ 
pair and the $B^+$ candidate, and $m_\pi$ is the pion mass.

The $B^+$ candidates are selected using independent artificial neural 
networks for each of the three $B^+$ decay modes.  The neural networks 
are based on the {\sc NeuroBayes} package \cite{neurobayes}.
For the decays $B^+\ra J/\psi K^+$ and $B^+\ra\bar{D}^0\pi^+$, we use 
the training and selection methods developed in Ref.~\cite{Aaltonen:2007tr}. 
For the decay $B^+\ra\bar{D}^0\pi^+\pi^+\pi^-$ we closely follow the
construction of the neural networks for the other two decays.
To train this last neural network we use data from the region 
$5325 < m(B^+) < 5395~\mevcc$ as the background sample and simulated $B^+$ 
events as the signal sample~\cite{MC}.
The most discriminating inputs to the neural networks are 
$p_T(B^+)$, $d_0(B^{+})$, $d_0$ of the kaon or pion with respect to 
the $B^+$ decay vertex, and the projected distance of the $B^+$ decay 
vertex from the primary vertex along the $B^+$ transverse momentum.
We select approximately $51~500$ $B$ events in
the $J/\psi K^+$ decay channel, $40~100$ in the $\bar{D}^0\pi^+$
channel, and $11~000$ in the $\bar{D}^0\pi^+\pi^+\pi^-$ channel.

To select $B^{**0}$ mesons, three additional neural networks are
trained on a combination of a simulated signal sample
and real data for a background sample. 
The data for the background sample are taken from the
entire $Q$ range of $0$ to $1000~\mevcc$,
which includes only a small contribution from the signal in the data.  
To avoid biasing 
the network training, the simulated events are
generated with the same $Q$ distribution as the data.
The $B^{**0}$ neural
networks use the same inputs as the $B^+$ neural networks,
together with the kinematic and particle 
identification quantities for the pion from the $B^{**0}$ decay.
The most important discriminants are the $p_T$ and $d_0$ of
the pion from the $B^{**0}$ decay vertex and
the output of the $B^+$ neural network.

\begin{figure}[htb]
  \centering
  \includegraphics[width=8.6cm]{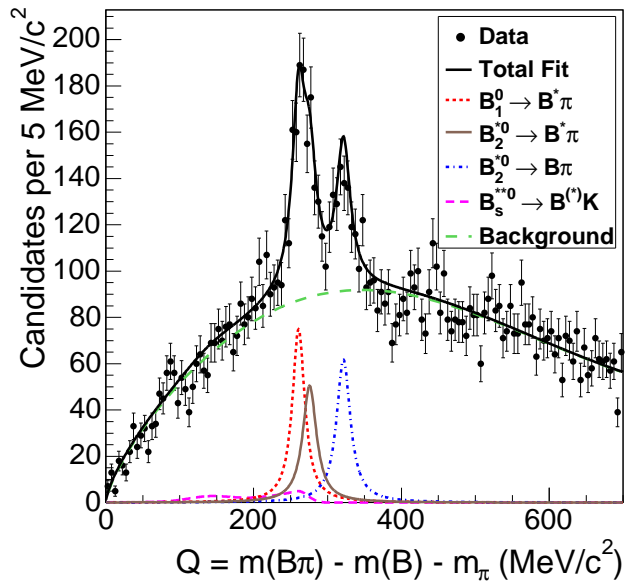}
  \parbox{0.8\linewidth}{
  \caption{Distribution of the mass difference
    	$Q = m(B^+\pi^-)-m(B^+)-m_\pi$ for exclusive $B^+$ decays.
	Curves are shown separately for the background, the $B_s^{**0}\to B^{(*)} K$ 
	reflections, and the three $B^{**0}$ decays.}
  \label{fig:baselineFit}}
\end{figure}

For each $B^+$ decay channel, we require fewer than six $B^{**0}$
candidates in an event in order to enhance the signal-to-background ratio.
The observed $B^{**0}$ signals are consistent for all three 
$B^+$ decay channels.  Therefore we combine the $B^{**0}$ events 
for all decay channels and use this combined $Q$ distribution to 
measure the $B^{**0}$ properties.  
We count the number of Monte Carlo
signal events, $N_{MC}$, and the number of signal and background events in 
the data, $N_{data}$, in the $Q$ signal region
of $200$ to $400~\mevcc$ for a given cut on each of the three network outputs.
We then optimize the $B^{**0}$ selection for each
$B^+$ decay channel to maximize the 
combined significance, $N_{MC}/\sqrt{N_{data}}$.
The resulting combined $Q$ distribution is shown in Figure 
\ref{fig:baselineFit}.

The $B^{**0}$ signal structure is interpreted
as resulting from the three signal processes
$B^0_1 \to B^{*+}\pi^-$, $B^{*0}_2 \to B^+\pi^-$, and 
$B^{*0}_2 \to B^{*+}\pi^-$, with $B^{*+}\rightarrow B^+\gamma$.  
The $Q$ distribution for each signal process is
modeled by a non-relativistic fixed-width Breit-Wigner 
function convoluted with the detector resolution model.
The resolution on Q is determined from simulation
and modeled as a sum of two Gaussian distributions,
a dominant narrow core and a broad tail with $Q$-dependent standard 
deviations of about $2~\mevcc$ and $4~\mevcc$, respectively.
The fraction of events in the broad tail is fixed to be 0.2.

We perform an unbinned maximum-likelihood fit to the combined
$Q$ distribution, from which we extract the $Q$ value of
the $B^{*0}_2 \to B^+\pi^-$ decay, the mass difference between the
$B^0_1$ and $B^{*0}_2$ states, the width of the $B^{*0}_2$, and the number 
of events in each signal process.  The following parameters in the fit are
constrained to their values from either
previous measurements or theoretical predictions:
the energy of the $B^{*+}$ decay photon,
$E(\gamma) = 45.78 \pm 0.35$ MeV/$c^2$ \cite{pdg};
the ratio of the $B^0_1$ and $B^{*0}_2$ widths, 
$\frac{\Gamma(B^0_1)}{\Gamma(B^{*0}_2)} = 0.9 \pm 0.2$~\cite{Falk:1995th};  
and the ratio of the $B^{*0}_2$ branching fractions, 
$\frac{BR(B^{*0}_2 \to B^+\pi^-)}{BR(B^{*0}_2 \to B^{*+}\pi^-)} = 1.1 \pm 0.3$
\cite{delphi_bds_ichep},
consistent with the value measured in Ref.~\cite{Abazov:2007vq}.

The background is modeled by a sum of two components, each being
the product of a power law and an exponential function.
We also expect reflections from $B_s^{**0} \to B^+ K^-$ decays 
when the kaon is mistakenly assigned the pion mass.  The shape of 
the reflection in the $Q$ distribution is determined in simulations 
of $B_s^{**0}$ states~\cite{Aaltonen:2007tr} and fixed in the fit.
The normalization of the $B_s^{**0}$ is obtained by correcting the observed
yield from Ref.~\cite{Aaltonen:2007tr} by a ratio of efficiencies
to reconstruct a $B_s^{**0}$ decay as a $B^{**0}$ and $B_s^{**0}$.
In the $B^{**0}$ data sample we expect $24 \pm 12$ $B^0_{s1}$ events and 
$62 \pm 31$ $B^{*0}_{s2}$ events.  These normalizations enter the fit 
as Gaussian constraints.

Sources of systematic uncertainty on the mass difference and width 
measurements include mass scale,
mass-dependent signal efficiency, fit model bias,
assumptions entered as Gaussian 
constraints in the fit, choice of background and 
resolution models, and location and amount of $B^{**0}$ broad states.
The systematic uncertainties are summarized in Table~\ref{tab:syst_summary}.

To determine the mass scale uncertainty, we reconstruct $\psi(2S) \to J/\psi\pi^+\pi^-$ 
with $J/\psi \to \mu^+\mu^-$, which has a similar
$Q$ value as the $B^{**0}$ decays.  We compare the measured $Q$
to the world average~\cite{pdg} and take the
difference as the mass scale uncertainty.
To evaluate the effect of signal efficiency with changing $Q$,
we generate a large number of samples of the same size 
as the data, called pseudoexperiments,
with the $Q$-dependent efficiency obtained from simulation.
We then apply the default fit to the pseudoexperiments.

Tests of the fit model on pseudoexperiments show a 
small fit bias on the $B^{**0}$ signal parameters, which is 
included as a systematic uncertainty.  Signal parameters entered 
as Gaussian constraints in the fit contribute to the fit 
uncertainty.  To determine their systematic contribution, we 
refit the data with these constrained parameters fixed.  This 
fit returns the statistical fit uncertainties, which are subtracted
in quadrature from the total fit uncertainties to obtain the 
systematic contribution.

To estimate the uncertainties due to the choice of background 
and resolution models, we generate pseudoexperiments with
varied background parameterizations or worse
mass resolution.
The background is also well-modeled by the sum of a
broad Breit-Wigner function with the product of a power law
and an exponential function.
From comparisons of the detector resolution in data and
Monte Carlo for the $\psi(2S)$ sample, we expect the Monte
Carlo to underestimate the resolution by no more than 20\%.
These pseudoexperiments are fit with the default fit and
the generating model.  The distribution of the differences 
between these fit results is modeled by a Gaussian, whose 
mean is assigned as the systematic uncertainty.

Possible effects of $B_0^*$ and $B_1^\prime$ decays on our 
background model are studied by adding two Breit-Wigner functions 
of identical width varied over the range $100-200~\mevcc$.
The $Q$ values of the states are independently varied in
the range 240 to 360 $\mevcc$, the region around the narrow
$B^{**0}$ peaks.  We refit the data for various masses and
widths of the broad states, with the normalizations of the 
broad Breit-Wigner functions as additional free parameters in the fit model.  
We then take the largest variation in the narrow $B^{**0}$ parameters from 
any configuration of broad states as the systematic uncertainty 
due to the $B^{**0}$ broad states.

\renewcommand{\arraystretch}{1.4} 
\begin{table}
\begin{center}
\caption{Systematic uncertainties on the $B^{**0}$ parameter measurements.
Each row corresponds to one source of systematic uncertainty.  The columns show
the uncertainties for each of the three $B^{**0}$ signal parameters.  
Uncertainties are in units of $\mevcc$.}
\label{tab:syst_summary}
\begin{tabular}{llll} 
\hline\hline
Source		& $Q(B^{*0}_2)$ 	& $\Gamma(B^{*0}_2)$    & $m(B^{*0}_2)-m(B^0_1)$ \\
\hline
Mass scale 	& $\pm0.2$             & --	          	& $<0.01$	 	\\
Efficiency      & $^{+0}_{-0.03}$       & $^{+0.4}_{-0}$      	& $^{+0}_{-0.3}$       \\
Fit bias	& $^{+0}_{-0.1}$	& $^{+0.4}_{-0}$      	& $^{+0}_{-0.2}$       \\
Fit constraints	& $^{+0.4}_{-0.3}$	& $^{+2.1}_{-1.5}$	& $^{+0.7}_{-0.9}$   	\\
Background	& $^{+0.2}_{-0}$	& $^{+0}_{-1.6}$	& $^{+0.2}_{-0}$	\\
Resolution	& $<0.01$	        & $^{+0}_{-0.4}$    	& $<0.01$		\\
Broad states	& $^{+0.7}_{-0.5}$	& $^{+2.3}_{-9.9}$	& $^{+0.9}_{-1.0}$	\\
\hline
Total		& $^{+0.9}_{-0.7}$	& $^{+3.2}_{-10.2}$	& $^{+1.2}_{-1.4}$	\\
\hline\hline
\end{tabular}
\end{center}
\end{table}
\renewcommand{\arraystretch}{1.0}

The result of the likelihood fit to the data is shown in 
Figure \ref{fig:baselineFit}, and we measure the following:
\begin{itemize}
\item[] $m(B^{*0}_2)-m(B^+)-m_\pi = 321.5^{+1.7}_{-1.8}$(stat.)$^{+0.9}_{-0.7}$(syst.) MeV/$c^2$;
\item[] $m(B^{*0}_2)-m(B^0_1) = 14.9^{+2.2}_{-2.5}$(stat.)$^{+1.2}_{-1.4}$(syst.) MeV/$c^2$; and
\item[] $\Gamma(B^{*0}_2) = 22.7^{+3.8}_{-3.2}$(stat.)$^{+3.2}_{-10.2}$(syst.) MeV/$c^2$.
\end{itemize}
The signal is consistent with theoretical predictions
\cite{Ebert:1997nk,Matsuki:2007zz},
and Gaussian-constrained parameters remain close to their input values, 
the largest departure being 0.4 standard deviations.
The numbers of events are
$N(B^0_1) = 503^{+75}_{-68}$, $N(B^{*0}_2\to B^+\pi^-) = 385^{+48}_{-45}$, 
and $N(B^{*0}_2\to B^{*+}\pi^-) = 351^{+48}_{-45}$, where uncertainties
are statistical only.
Using the mass of the $B^+$ \cite{pdg} and the correlations between the
fit parameters, the masses of the $B^0_1$ and $B^{*0}_2$ are
$m(B^{*0}_2) = 5740.2^{+1.7}_{-1.8}$(stat.)$^{+0.9}_{-0.8}$(syst.) MeV/$c^2$ and
$m(B^0_1) = 5725.3^{+1.6}_{-2.2}$(stat.)$^{+1.4}_{-1.5}$(syst.) MeV/$c^2$.
With the current statistics the data are also consistent 
with containing only the $B^0_1$ and $B^{*0}_2 \to B^+\pi^-$ peaks.

In summary, using the three fully reconstructed decays 
$B^+ \to J/\psi K^+$, $B^+ \to \bar{D}^0 \pi^+$, and
$B^+ \to \bar{D}^0 \pi^+ \pi^+ \pi^-$, we observe two
narrow $B^{**0}$ states in the decays $B^0_1 \to B^{*+} \pi^-$
and $B^{*0}_2 \to B^{(*)+} \pi^-$.  This is the most precise 
measurement of the narrow $B^{**0}$ masses to date.
We have also measured the $B^{*0}_2$ width for the first time.
There is some discrepancy between these measurements 
and those reported by the D0~collaboration~\cite{Abazov:2007vq},
the largest being close to a $3~\sigma$ difference in the mass 
splitting of the two $B^{**0}$ states. 

We thank the Fermilab staff and the technical staffs of the 
participating institutions for their vital contributions. 
This work was supported by the U.S. Department of Energy and 
National Science Foundation; the Italian Istituto Nazionale 
di Fisica Nucleare; the Ministry of Education, Culture, 
Sports, Science and Technology of Japan; the Natural 
Sciences and Engineering Research Council of Canada; the 
National Science Council of the Republic of China; the Swiss 
National Science Foundation; the A.P. Sloan Foundation; the 
Bundesministerium f\"ur Bildung und Forschung, Germany; the 
Korean Science and Engineering Foundation and the Korean 
Research Foundation; the Science and Technology Facilities 
Council and the Royal Society, UK; the Institut National de 
Physique Nucleaire et Physique des Particules/CNRS; the 
Russian Foundation for Basic Research; the Ministerio de 
Ciencia e Innovaci\'{o}n, Spain; the Slovak R\&D Agency; 
and the Academy of Finland. 



\end{document}